\documentclass[sn-mathphys]{sn-jnl}
\jyear{2021}
\usepackage{float}

\begin{document}

\title[High Density Electrons Above a Helium Film]{Measurement of High Density Electrons Above a Helium Film on an Amorphous Metal Substrate}

\author{\fnm{K. E.} \sur{Castoria}}\email{castoria@princeton.edu}

\author{\fnm{S. A.} \sur{Lyon}}\email{lyon@princeton.edu}

\affil{\orgdiv{Department of Electrical Engineering}, \orgname{Princeton University}, \orgaddress{\city{Princeton}, \postcode{08544}, \state{New Jersey}}}


\abstract{We have measured two-dimensional electron systems bound to a thin helium film supported by a metallic substrate. We report on our measurement of electron density obtained via a Kelvin probe technique. The underlying metallic substrate is an amorphous metallic  alloy (TaWSi), which can support large uniform densities due to its low surface roughness and homogeneous work function. We find that this substrate is able to support high enough densities that the electrons are expected to be Fermi-degenerate.  }


\maketitle

\section{Introduction}\label{Intro}

Electrons bound to the surface of superfluid helium form a very clean 2D electron gas owing to the smooth helium surface and the electrons’ relative isolation inside the  vacuum\cite{Monarkha_2004}. The system offers a unique platform from which to probe new regimes in two-dimensional physics. One area of particular interest is understanding Wigner crystallization with Fermi degenerate electrons. For over forty years, quantum melting of a Wigner crystal has been an elusive experimental goal driven by the predictions of exotic phases\cite{Spivak_2010} and quantum magnetism\cite{Ortiz_1999}. The phase boundary separating the solid electron phase from the fluid phase goes up to 2.4$\times$ 10$^{12}$ cm$^{-2}$ for electrons on bulk helium at T = 0 K \cite{Peeters_1983}, a density for which the electron system is degenerate. For bulk helium however, this is experimentally infeasible, as a hydrodynamic instability prevents the accumulation of electron densities higher than 2.2$\times$10$^9$ cm$^{-2}$\cite{Williams_1971}. On thin films of helium however, the maximum density that can be supported increases as the film gets thinner\cite{Peeters_1984}. Films on the order of ten nanometers and less can support densities in the range of a 10$^{11}$ – 10$^{12}$ cm$^{-2}$\cite{Lytvynenko_2017}. In this paper, we report measurements up to an electron density $n = 4.74\times 10^{11}$ cm$^{-2}$ on a helium film calculated to be 2 nm thick.

\begin{figure}[H]%
\centering
\includegraphics[width=0.9\textwidth]{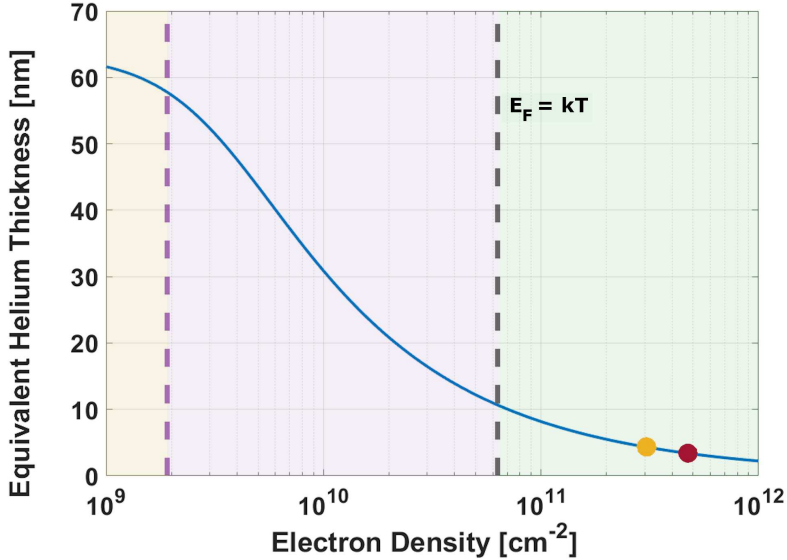}
\caption{Calculated Equivalent Helium Thickness (as defined in the text) versus electron density (blue). As the density increases and the thickness decreases, the electron system transitions from a classical gas (yellow region) to a classical Wigner solid (purple region). The classical melting point of the Wigner crystal as calculated by Saitoh\cite{Saitoh_1989} is shown as the purple dashed line. The density for which the system's Fermi energy is equal to our operating temperature of 1.75 K is denoted by the black dashed line. For densities above this point, the system is Fermi degenerate (green region). Also plotted are our measured densities discussed later in the text (yellow and red points).  }
\label{PhaseDiagram}
\end{figure}

If the substrate below the helium film is metallic, the electron-electron interaction is significantly altered. As the inter-electron spacing becomes larger than the distance between an electron and its image, the interaction is largely screened by the substrate. This effect depends on a quantity referred to in Fig.~\ref{PhaseDiagram} as Equivalent Helium Thickness (EHT), which is defined as 
\begin{equation}
EHT = d + r_e \epsilon_{He}
\label{eq:EHT}
\end{equation}
Where $d$ is the helium film thickness, $r_e$ is the distance between the electron layer and helium film surface, and $\epsilon_{He}$ is the dielectric constant of helium. 

Most notably the screening opens a gap in the phase diagram such that even as the temperature approaches zero, the system remains a liquid for finite densities. The blue curve in Fig.~\ref{PhaseDiagram} shows the EHT as a function of density for a helium film supported by our metal plate 2 mm above the bulk helium surface. The details of this calculation are discussed in Sec.~\ref{DescExp}. The classical Wigner crystal melting point as calculated by Saitoh\cite{Saitoh_1989} is marked with a purple dashed line. For densities lower than this value, the system is a classical electron gas, and for densities larger than this value, the system is a Wigner solid. As the density is increased, the helium film is compressed and the electron is bound closer to the helium surface, increasing the screening. It is expected that on helium films of ten nanometers and less, the Wigner crystal is no longer stable at electron densities on the order of 10$^{11}$ cm$^{-2}$. At these densities, the Fermi energy is well above our experimental temperature of 1.75 K (the black dashed line). While classical Wigner transitions have been well studied both experimentally and theoretically\cite{grimes_1979,Kosterlitz_1973}, the processes governing melting in the Fermi degenerate regime are not yet fully known. Both the possibility of polaron formation and quantum effects can significantly change the phase diagram in the quantum regime. 

Observing quantum melting has been difficult in part due to substrate disorder, which leads to both electron loss and reduced mobility, precluding transport measurements. We believe these problems can be mitigated with the use of amorphous metals. Our helium film in this experiment is supported by a very smooth amorphous metal (TaWSi). While this substrate is expected to allow transport measurements\citep{Asfaw_2019}, none are reported here. Further experiments will be required to fully understand the phase of the electron system present. In this work, a description of the experimental set up and the measurement techniques are given, followed by details about the amorphous metal substrate and concluding with a report of our measured electron density. 

\section{Description of Experiment}\label{DescExp}
\begin{figure}[H]%
\centering
\includegraphics[width=0.9\textwidth]{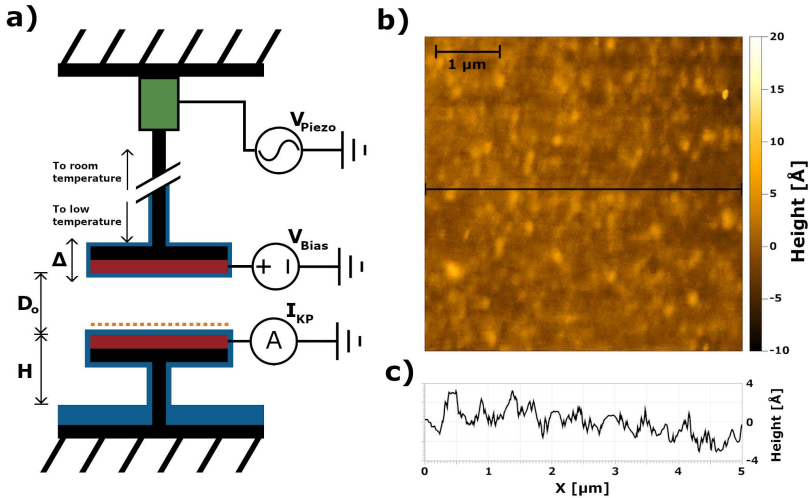}
\caption{\textbf{a} Schematic of the experimental setup for measuring the electron density. The electrons (orange dots) are bound to the helium film (blue). The film is supported by the lower plate (red) which is a height, $H$, above the bulk helium level and connected to a lock-in amplifier to measure the Kelvin probe current ($I_{KP}$). The upper plate (red) is a distance, $D_0$, above the lower plate and connected to a DC voltage source, $V_{Bias}$. This plate moves vertically by an amount, $\Delta$, as determined by the AC drive voltage, $V_{Piezo}$, applied to the the piezo stack (green). \textbf{b} The surface profile of a 5 $\mu$m$\times$5 $\mu$m patch of TaWSi sputtered on sapphire measured via atomic force microscopy. \textbf{c} The height profile along the black line pictured in \textbf{b}. }
\label{Schematic}
\end{figure}

The results presented in this paper are measured with a Kelvin probe technique\cite{Leners_1972} that can determine electron density irrespective of the electron mobility. This is done so that measurements of electron density can be made even if the electrons are localized due to crystallization, disorder, or polaron formation. The basic experimental system is illustrated in Fig.~\ref{Schematic}a. A small controllable amount of helium is let into our leak tight cell until a desired quantity of helium is condensed at the bottom. Below 2.2 K, the helium becomes a superfluid and a thin van der Waals film coats all the surfaces in the cell above the bulk level\cite{Atkins_1950,Sabisky_1973}. At a height, $H$, above this inner liquid level, a metal plate is mounted that supports the thin film and 2D electrons. Together with a second metal plate mounted a height, $D_0$, above the first, a simple parallel plate capacitor is formed. This capacitor is the Kelvin probe that will be used to measure electron density. The upper plate has a small hole cut out, through which a tungsten filament can emit electrons onto the lower helium surface. The filament and upper plate are mounted on a stainless-steel tube that runs to room temperature and terminates on a small piezo stack to drive vertical motion.

The top and bottom capacitor plates are made of an amorphous metallic alloy. We are using Ta$_{40}$W$_{40}$Si$_{20}$ (referred to as TaWSi), which has been used previously in experiments showing transport of electrons on helium in channel devices with helium films down to 200 nm \cite{Asfaw_2019}. These transport results suggest that TaWSi exhibits the low surface potential disorder required to support a uniform electron density across the whole Kelvin probe area. The bottom plate consists of 100 nm of TaWSi sputtered onto a polished sapphire substrate 2.25 cm$^2$ in area. The top plate consists of 100 nm of TaWSi sputtered onto the copper pad of a printed circuit board. Having both the top and bottom plate made of TaWSi simplifies Eq.~\ref{eq:IKP} by setting $\Delta_\Phi = 0$.  Fig.~\ref{Schematic}b shows the surface topography for a 5 $\mu$m $\times$ 5 $\mu$m square of a typical TaWSi film sputtered on polished sapphire as measured by an atomic force microscope, and the profile along the center line of this area is shown in Fig.~\ref{Schematic}c. The RMS surface roughness is measured to be 1.5 \AA .

	The piezo stack can move the top plate by a maximum distance, $\Delta$, and is controlled with an AC voltage, $V_{Piezo}$. By driving the stack  at a frequency, $\omega$, the top plate displacement creates a time-varying capacitance. An AC current, $I_{KP}$, at the drive frequency, $\omega$, will then flow given a DC voltage across the capacitor. In our experiment the DC voltage across the capacitor depends on an externally applied voltage difference, $V_{Bias}$, the work function difference between the top and bottom metals, $\Delta_\Phi$, and the electron potential of the 2D electrons on the helium film. By differentiating the equation for the charge on a capacitor with respect to time, we find that the current through the capacitor is 
	
\begin{equation} 
I_{KP} = \frac{\partial}{\partial t}\left(\frac{\epsilon_0 A}{D(t)} \right) \left( V_{Bias}+V_e  + \Delta_\Phi \right)
\label{eq:IKP_total}
\end{equation}		
	
where $\epsilon_0$ is the permitivity of free space, $A$ is the plate area, $D(t)$ is the time dependent plate separation distance, and $V_e$ is the electron potential.
	
We drive the piezo such that $D(t) = D_0 + \Delta sin \left( \omega t \right)$. In the limit $\left(\frac{\Delta}{D_0} \ll 1\right)$, it can be shown that the current through the capacitor at the drive frequency is given by   	
	
\begin{equation} 
I_{KP} = C_0 \omega \frac{\Delta_{RMS}}{D_0}  \left( V_{Bias} + \Delta_\Phi + V_e \right)
\label{eq:IKP}
\end{equation}

where $C_0$ is the mean parallel plate capacitance and $\Delta_{RMS}$ is the RMS travel of the top plate.

From Eq.~\ref{eq:IKP}, it can be seen that by sweeping $V_{Bias}$ and using a lock-in amplifier to measure $I_{KP}$ at $\omega$, we expect to measure a linear I-V curve. Further, by charging the helium film and sweeping $V_{Bias}$ again, we expect the line to be shifted by a voltage offset proportional to $V_e$. 

We then need to to determine the electron density from the measured electron potential. The helium film thickness, $d$, is much smaller than the mean plate separation, $D_0$, so we can ignore the electrons' capacitance to the top plate and just consider their potential from their capacitance to the bottom plate. The film thickness for densities above 10$\times$10$^{11}$ cm$^{-2}$ is on the order of single nanometers, comparable to the size of the electrons' wavefunction. For this reason, it is necessary to numerically solve the Schrodinger equation to account for the wavefunction extent of the electron in computing its capacitance to the bottom plate\cite{Joseph_2021}. 

For a given electron density and substrate height above the bulk helium, the film thickness on a plate is well understood\cite{Sabisky_1973,Leiderer_1984}. Using this thickness, we compute the electron wavefunction normal to the surface. We use this wavefunction to find $V_e$ by integrating the potential from the bottom plate and find

\begin{equation} 
V_e = \int_{0}^{\infty}  \psi^\dagger(z)\psi(z)\frac{4\pi n e}{\epsilon_0 \epsilon_{He}} \left( d + z \epsilon_{He} \right)  dz
\label{eq:electronPotential}
\end{equation}	

where $e$ is the electronic charge and $z$ is the coordinate normal to the helium surface. This calculation implicitly assumes that the TaWSi is a perfect metal. From spectroscopic ellipsometry of the TaWSi we estimate a plasma energy of about 10 eV which would be consistent with this assumption.

Using this, we get a one-to-one mapping between electron density and electron potential. We therefore can use the measured voltage shifts of the Kelvin probe I-V curve to determine the electron density and film thickness.

\section{TaWSi Plate Results}\label{kelvinResults}

\begin{figure}[H]%
\centering
\includegraphics[width=0.9\textwidth]{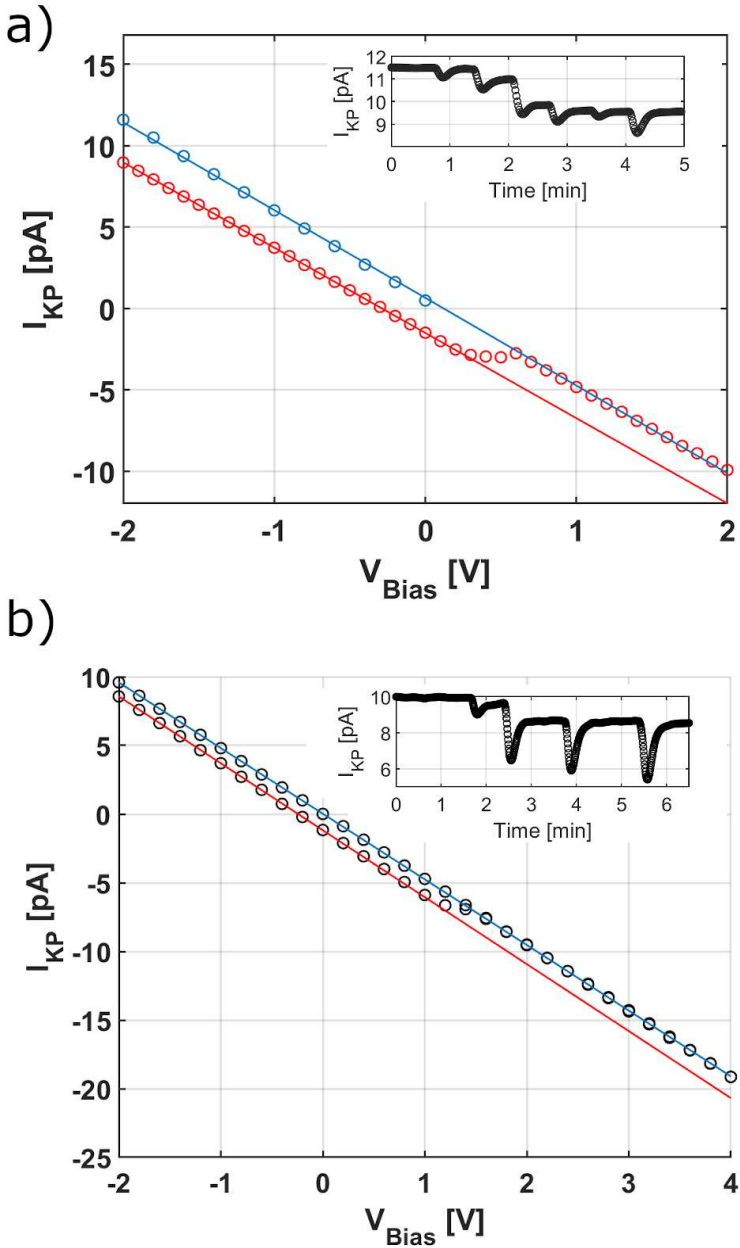}
\caption{ Kelvin probe measurements of the 2D electron system. \textbf{a} Measurement of the Kelvin probe current before (blue circles) and after (red circles) while scanning the top plate voltage, $V_{bias}$. Linear fits to both sweeps over the region $V_{Bias} < 0$ are shown as solid lines. \textbf{b} Measurement of the Kelvin probe current while sweeping $V_{Bias}$ from -2 V to 4 V and back to -2V. Overlaid in red (blue) is a linear fit to the I-V curve while the film is charged (uncharged). Both insets show the charging of the helium surface as a function of time with discrete filament pulses about once per minute, while $V_{Bias}$ is held at -2 V.} 
\label{BiasShiftPlot}
\end{figure}

Results from this Kelvin probe apparatus show that we were able to charge our helium films to densities above 10$^{11}$ cm$^{-2}$. Measurements from two such experiments are shown in Fig.~\ref{BiasShiftPlot} with the insets showing the Kelvin probe current shifting during charging. For the results presented in Fig.~\ref{BiasShiftPlot}a, the piezo is driven its full travel range of $\Delta_{RMS}$ = 17 $\mu$m at $\omega$ = 14.5 Hz. The bottom plate is mounted about 2 mm above the bulk helium and the plates are separated by D$_0$ = 404 $\mu$m. Prior to emission, $V_{Bias}$ is swept from -2 V to 0 V and the current through the uncharged probe is recorded (blue circles). With the top plate held at $V_{Bias} = -2V$ and no electrons emitted, the uncharged current through the probe is measured to be 11.68 pA. This is slightly less than the expected current of 13.5 pA, possibly due to the imperfect AC response of the piezo. The tungsten filament is pulsed until no more change in the Kelvin probe current is observed with further flashes. A typical charging plot is shown in the inset. Once charged, $V_{Bias}$ is again swept, this time to +2 V (red circles). For $V_{Bias} < 0$, the measured current follows a shifted I-V curve (red line). Once the top plate voltage becomes more positive than ~0.5 V, the electrons desorb from the helium film, and the measured current closely follows the linear fit to the uncharged I-V curve (blue line). The voltage offset between these fits is 300 mV, which corresponds to a density of 4.74$\times$10$^{11}$ cm$^{-2}$, film thickness of 2.0 nm, and an EHT of 3.4 nm. As long as $V_{Bias}$ is kept at the initial -2 V, the electrons are stably held on the surface. No change in density was observed while held for a period of thirty minutes. 

In Fig.~\ref{BiasShiftPlot}b, a similar experiment is shown. In this experiment, the drive frequency was 17.5 Hz and the plates were measured to be 483 $\mu$m apart. Here, $V_{Bias}$ is swept from -2 V to 4 V and back, all after emitting electrons (as shown in the inset). The electrons desorbed once $V_{Bias}$ became more positive than 1.4 V. Until this point, the I-V curve closely followed the charged fit (red line). For the remainder of the sweep to 4 V and the entire sweep back to -2 V, the electrons followed the uncharged fit (blue line). A voltage shift of 247 mV is measured, which corresponds to a density, film thickness, and EHT of 3.06$\times$10$^{11}$ cm$^{-2}$, 2.7 nm, and 4.3 nm respectively. 

\section{Channel Device Results}\label{ST}

\begin{figure}[H]%
\centering
\includegraphics[width=0.9\textwidth]{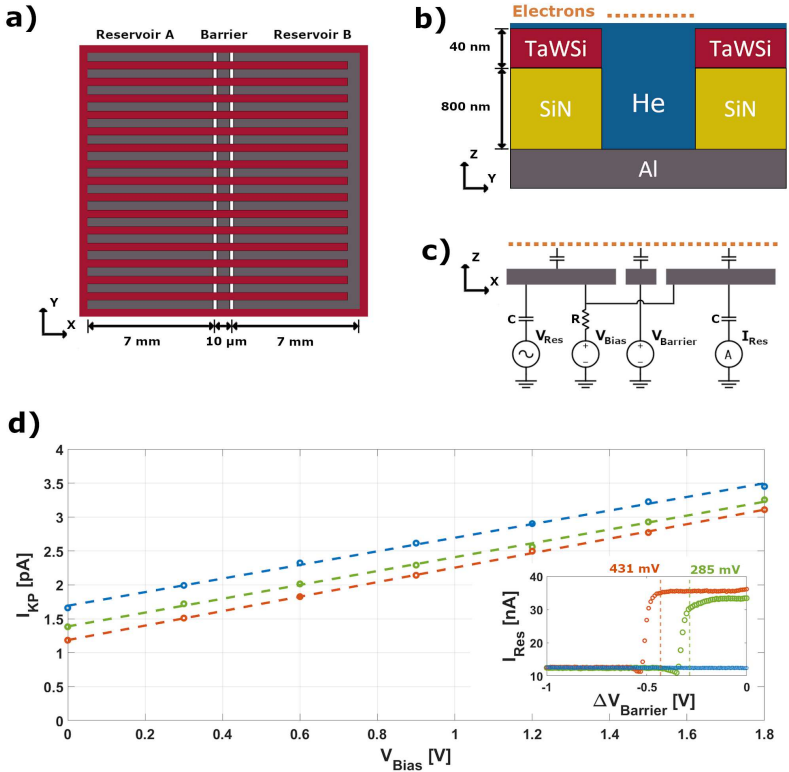}
\caption{Two layer channel device diagrams and measurements \textbf{a} Top down view of the channel device. The 1210 channels are 14 mm long, 7 $\mu$m wide, and separated by 7 $\mu$m wide walls. \textbf{b} Side view of the channel device. The channels are comprised of 800 nm of silicon nitride (yellow) and 40 nm of TaWSi (red) deposited onto the bottom aluminum layer (grey). Superfluid helium (blue) fills the channels due to capillary action. \textbf{c} Circuit diagram of the transport measurements. In all of the experiments discussed, the upper TaWSi layer is tied to ground. The values $R$ and $C$ are chosen to have no effect in the frequency range of operation. \textbf{d} Measurements of the Kelvin probe current (as measured from the top capacitor plate) while sweeping the bias on the aluminum plates. The scans are taken when the  channels are empty (blue), partially filled (green), and full (orange). The inset shows corresponding transport current along the channels while sweeping the barrier gate. The dashed vertical lines show the voltage offsets measured by the Kelvin probe.} 
\label{SommerTannerPlot}
\end{figure}

The Kelvin probe shifts are susceptible to stray capacitance with the environment\cite{Baikie_1991}, so it is conceivable that charging insulating surfaces inside the cell or helium films outside of the capacitor region could lead to misleading voltage shifts. To confirm that the electrons on the bottom plate's film cause the observed shifts, a similar Kelvin probe experiment is performed as described in Sec.~\ref{kelvinResults}. In these experiments, the bottom TaWSi plate is replaced with a two layer device consisting of helium filled channels of the same area, and $V_{Bias}$ is connected to two large area gates on this device, as shown in Fig.~\ref{SommerTannerPlot}. The top plate is biased to a negative DC voltage (-2.5 V in the experiments shown) and capacitively coupled to the lock-in amplifier to measure the $I_{KP}$. 

A top down diagram of the two layer device is shown in Fig~\ref{SommerTannerPlot}a. The bottom layer (grey) is made of aluminum and is divided into three gates. The large outer two reservoir gates are 7 $\times$ 18 mm$^2$ and the center barrier gate is 0.01 $\times$ 18 mm$^2$. The top layer of TaWSi (red) divides these gates into 1210 channels, each 7 $\mu$m wide. A cross section of one such channel filled with helium is shown in Fig~\ref{SommerTannerPlot}b. An 800 nm layer of silicon nitride (yellow) below the 40 nm thick TaWSi creates channels 840 nm deep. With the device above the bulk helium, the channels are filled with superfluid helium by capillary action\cite{Marty_1986}. Electrons are confined to the channels by biasing the bottom aluminum gates positive relative to the TaWSi layer at the top of the channels.

To determine electron potential in the channels, we measure electron transport across the device as a function of the barrier gate voltage. A circuit diagram of this scheme is shown in Fig~\ref{SommerTannerPlot}c. Both reservoirs are connected to a positive DC bias. One of the reservoirs is capacitively coupled to an AC voltage, $V_{Res}$, with a frequency $\omega_{Res} >> \frac{1}{RC}$, and the other is coupled to a lock-in amplifier to measure the induced current, $I_{Res}$. Most of the current is carried by the surface electrons, so it can be controlled by the difference between the barrier gate voltage and the reservoirs' DC bias $\Delta V_{Barrier}$. By sweeping $\Delta V_{barrier}$ negative, the current will remain constant until the barrier potential is equal to the electron potential in the channels. As it becomes more negative than the electron potential, the electrons in the channels can no longer travel between reservoirs, and a sharp decrease in $I_{Res}$ is observed. The voltage where this transition occurs is equal to the electron potential in the channels. 

To show that the voltage shift measured by the Kelvin probe is only due to electrons above the bottom capacitor surface, the electron potential is measured both via channel transport and the Kelvin probe. The channels are charged to two different densities as shown in Fig.~\ref{SommerTannerPlot}d. Measurements are shown with the channels empty (blue), partially filled with electrons (green), and charged to saturation (orange). The channel transport measurements are shown in the inset with dashed lines indicating the corresponding Kelvin probe voltage shift (285 mV for the low density and 431 mV for the high density). In both cases, the shift voltage is in good agreement with the $\Delta V_{barrier}$ where transport begins to turn off.

\section{Conclusion}\label{Conclusion}

In these measurements, we have shown that high densities of electrons can be stably held on a thin helium surface over a metallic substrate. Using amorphous metals and fabricating smoother substrates, we are able to hold higher densities than previously reported in this system. These densities were determined with a Kelvin probe measurement technique.  With densities above 4$\times$10$^{11}$ cm$^{-2}$, the Fermi energy is an order of magnitude larger than the ambient temperature of 1.75K. Using computed electron wavefunctions, electron densities and helium film thicknesses were determined from the measured voltage shifts. The results of this measurement technique were compared to an independent measurement of electron density via transport in channels. We find that the two different methods are in good agreement with each other. In order to get information regarding the nature of the many-body electron system, further measurements will be necessary. The calculations of Peeters and Platzman\cite{Peeters_1983} suggest that the Wigner crystal will not form at these densities and film thicknesses. Unless the electron mass is increased, due to polaron formation for example, the electrons are Fermi degenerate. However, it is thought that the the self-trapping energy for polarons is on the millikelvin scale\cite{Jackson_1986}, well below the temperatures used here and far below the Fermi temperature. Disorder may also impede the formation of a Fermi liquid. In some situations it is thought that isolated "puddles" of a Fermi liquid are formed, but disorder does not allow them to merge\cite{Deng_2016}. The strength of the disorder seen by these electrons is not yet known.

\section{Data Availability}\label{Data Availability}

The datasets generated during and/or analysed during the current study are available from the corresponding author on reasonable request.

\section{Acknowledgments}\label{Acknowledgments}
This material is based upon work supported by the U.S. Department of Energy, Office of Science, Office of Basic Energy Sciences, under Award number DE-SC0020136. Support was provided, in part, through the Program in Plasma Science and Technology under DOE Contract Number DE-AC02-09CH11466.

\section{Compliance with Ethical Standards}\label{Compliance with Ethical Standards}
S.A Lyon is an officer of Eeroq Corp. 

\clearpage

\nocite{label}

\bibliography{MyBib}

\end{document}